\PassOptionsToPackage{unicode}{hyperref}
\PassOptionsToPackage{hyphens}{url}
\documentclass[
]{article}
\usepackage{xcolor}
\usepackage[margin=1in]{geometry}
\usepackage{amsmath,amssymb}
\usepackage{iftex}
\ifPDFTeX
  \usepackage[T1]{fontenc}
  \usepackage[utf8]{inputenc}
  \usepackage{textcomp} 
\else 
  \usepackage{unicode-math} 
  \defaultfontfeatures{Scale=MatchLowercase}
  \defaultfontfeatures[\rmfamily]{Ligatures=TeX,Scale=1}
\fi
\usepackage{lmodern}
\ifPDFTeX\else
\fi
\IfFileExists{upquote.sty}{\usepackage{upquote}}{}
\IfFileExists{microtype.sty}{
  \usepackage[]{microtype}
  \UseMicrotypeSet[protrusion]{basicmath} 
}{}
\makeatletter
\@ifundefined{KOMAClassName}{
  \IfFileExists{parskip.sty}{%
    \usepackage{parskip}
  }{
    \setlength{\parindent}{0pt}
    \setlength{\parskip}{6pt plus 2pt minus 1pt}}
}{
  \KOMAoptions{parskip=half}}
\makeatother
\usepackage{longtable,booktabs,array}
\usepackage{calc} 
\usepackage{etoolbox}
\makeatletter
\patchcmd\longtable{\par}{\if@noskipsec\mbox{}\fi\par}{}{}
\makeatother
\IfFileExists{footnotehyper.sty}{\usepackage{footnotehyper}}{\usepackage{footnote}}
\makesavenoteenv{longtable}
\usepackage{graphicx}
\makeatletter
\newsavebox\pandoc@box
\newcommand*\pandocbounded[1]{
  \sbox\pandoc@box{#1}%
  \Gscale@div\@tempa{\textheight}{\dimexpr\ht\pandoc@box+\dp\pandoc@box\relax}%
  \Gscale@div\@tempb{\linewidth}{\wd\pandoc@box}%
  \ifdim\@tempb\p@<\@tempa\p@\let\@tempa\@tempb\fi
  \ifdim\@tempa\p@<\p@\scalebox{\@tempa}{\usebox\pandoc@box}%
  \else\usebox{\pandoc@box}%
  \fi%
}
\def\fps@figure{htbp}
\makeatother
\NewDocumentCommand\citeproctext{}{}
\NewDocumentCommand\citeproc{mm}{%
  \begingroup\def\citeproctext{#2}\cite{#1}\endgroup}
\makeatletter
 \let\@cite@ofmt\@firstofone
 \def\@biblabel#1{}
 \def\@cite#1#2{{#1\if@tempswa , #2\fi}}
\makeatother
\newlength{\cslhangindent}
\newlength{\csllabelwidth}
\newenvironment{CSLReferences}[2] 
 {\begin{list}{}{%
  \setlength{\itemindent}{0pt}
  \setlength{\leftmargin}{0pt}
  \setlength{\parsep}{0pt}
  \ifodd #1
   \setlength{\leftmargin}{\cslhangindent}
   \setlength{\itemindent}{-1\cslhangindent}
  \fi
  \setlength{\itemsep}{#2\baselineskip}}}
 {\end{list}}
\usepackage{calc}

\providecommand{\tightlist}{%
  \setlength{\itemsep}{0pt}\setlength{\parskip}{0pt}}
\usepackage{setspace}
\usepackage{booktabs}
\usepackage{booktabs}
\usepackage{longtable}
\usepackage{array}
\usepackage{multirow}
\usepackage{wrapfig}
\usepackage{float}
\usepackage{colortbl}
\usepackage{pdflscape}
\usepackage{tabu}
\usepackage{threeparttable}
\usepackage{threeparttablex}
\usepackage[normalem]{ulem}
\usepackage{makecell}
\usepackage{xcolor}
\usepackage{bookmark}
\IfFileExists{xurl.sty}{\usepackage{xurl}}{} 
\hypersetup{
  pdftitle={Towards a valid bibliometric measure of epistemic breadth of researchers},
  pdfauthor={Paul Donner; Clemens Blümel},
  hidelinks,
  pdfcreator={LaTeX via pandoc}}

\title{Towards a valid bibliometric measure of epistemic breadth of researchers}
\author{Paul Donner\footnote{email: \href{mailto:donner@dzhw.eu}{\nolinkurl{donner@dzhw.eu}}. ORCID: 0000-0001-5737-8483. Department 2 `Research System and Science Dynamics', German Centre for Higher Education Research and Science Studies (DZHW), Berlin, Germany} \and Clemens Blümel\footnote{ORCID: 0000-0003-2465-6840. Department 2 `Research System and Science Dynamics', German Centre for Higher Education Research and Science Studies (DZHW), Berlin, Germany}}
\date{}

\begin{document}
\maketitle

\textbf{Abstract}\\
The concept of \emph{epistemic breadth} of the work of a researcher refers to the scope of their knowledge claims, as reflected in published research reports. Studies of epistemic breadth have been hampered by the lack of a validated measure of the concept. Here we introduce a knowledge space approach to the measurement of epistemic breadth and propose to use the semantic similarity network of an author's publication record to operationalize a measure. In this approach, each paper has its own location in a common abstract vector space based on its content. Proximity in knowledge space corresponds to thematic similarity of publications. Candidate measures of epistemic breadth derived from aggregate similarity values of researchers' bodies of work are tested against external validation data of researchers known to have made a major change in research topic and against self-citation data. We find that some candidate measures co-vary well with known epistemic breadth of researchers in the empirical data and can serve as valid indicators of the concept.


\section{Introduction}\label{introduction}

Researchers do not work on only a single research problem throughout their careers. They advance knowledge claims to problems in different topics, for various reasons and with differing frequencies (\citeproc{ref-gieryn1978problem}{Gieryn, 1978}; \citeproc{ref-glaser2015bibliometric}{Gläser \& Laudel, 2015}), sometimes even changing their major disciplines (\citeproc{ref-harmon1965profiles}{Harmon, 1965, pp. 50--52}; \citeproc{ref-lepair1980switching}{Le Pair, 1980}). The forms and patterns of these changes are manifold, as are the external factors influencing them. New topics may be more or less thematically related to previous topics. Expansions of research portfolios can be caused by preceding career changes or they may be the reason for career changes -- we don't know which is more often the case. Enders, Kehm, \& Schimank (\citeproc{ref-enders2014turning}{2014}) show that funding opportunities, strategic institutional priorities, and assessment results can shape research problem choice of research groups.

Research topic changes can also influence the social relations of researchers by opening opportunities for collaboration or by creating the need to seek out new colleagues for joint research projects (\citeproc{ref-tripodi2020knowledge}{Tripodi, Chiaromonte, \& Lillo, 2020}). There may also be field-specific incentives or barriers for thematic changes. For example, in some fields it is expected of scholars to change research topics in order to attain a tenured position as a professor (\citeproc{ref-bielick2018emergence}{Bielick \& Laudel, 2018}). One potential barrier to epistemic change is a perceived risk of reduced scientific productivity, as the engagement with novel fields of study requires an investment of time and resources with an uncertain outcome (\citeproc{ref-franzoni2017academic}{Franzoni \& Rossi-Lamastra, 2017}). In addition, formal regulations of, and perceived norms embedded in, evaluation systems, such as competitive grant funding review procedures or requirements for tenure, may affect scholars' research strategies in such a way as to diversify or widen their topical portfolio in order to manage uncertainties (\citeproc{ref-fochler2018anticipatory}{Fochler \& Sigl, 2018}; \citeproc{ref-muller2019re}{Müller \& Kaltenbrunner, 2019}). All this happens in a highly dynamic, evolving cognitive environment in which older narrow topics become epistemically exhausted and new ones are introduced incessantly.

Thus, the development of their thematic research profile is a central concern for researchers and it is a site where researchers' cognitive careers (e.g., the portfolio and sequence of topics on which a researcher works) and institutional careers, as the different positions researchers hold throughout their professional life (\citeproc{ref-bielick2018emergence}{Bielick \& Laudel, 2018}), interact. Scholarship in science studies has therefore investigated conditions, but also the causes and effects, of changes in both cognitive and institutional careers. Increasingly, science and innovation policy-makers are interested in establishing incentives for researchers to engage with novel fields combining different realms of knowledge in order to align research priorities and societal goals, thus it is timely to improve our understanding of the development of researchers' thematic profiles, more knowledge about the nature of such changes and their effects is needed. In particular, we argue, it is necessary to be able to validly measure and compare properties of the knowledge structures of individual scholars.

In this article, we aim to contribute to the literature by proposing a measure for `epistemic breadth'. We understand epistemic breadth as the extent to which researchers' research portfolios encompass multiple research subjects.

Despite the interest in researchers' knowledge structures and their dynamics, there is no common understanding of the concept, due to a plethora of competing terms and concepts closely related to it, such as topical switch, cognitive mobility, intellectual migration, research problem choice, and others.\footnote{We note that \emph{interdisciplinarity} of research is a closely related but distinct concept and many instructive insights can be gleaned from the lively discussion about its meaning and operationalization for measurement. A useful summary of this literature can be found in Rousseau, Zhang, \& Hu (\citeproc{ref-rousseau2019knowledge}{2019}). According to their characterization, the central element of interdisciplinarity is \emph{integration} of different knowledge stocks, or overcoming the disparity of extant knowledge. Integration is not a constitutive element of the concept of epistemic breadth: a researcher can exhibit an epistemically broad research portfolio by following different research strands which remain disparate. A very topically specialized researcher, on the other hand, can very well contribute to a highly interdisciplinary research project by bringing together her specialized knowledge and skills with complementary contributions by others.}

Our objective in this contribution is to provide a validated bibliometric measure for epistemic breadth. We operationalize this concept on the basis of cognitive structure. Specifically, we reason that the overall `breadth' of the thematic portfolio of a researcher's published knowledge claims is an emergent property of the network of \emph{epistemic distances} of a researcher's knowledge contributions in relation to a knowledge space. We posit that the semantic distance between the epistemic contributions of an individual researcher is a relevant aspect that has been overlooked by scholarship that often focussed on the frequency of topic changes or the sequence of topics. While there have been multiple prior proposals for indicators tapping constructs similar or equivalent to epistemic breadth, our approach differs from these in not relying on pre-specified classification systems or \emph{ad hoc} clustering routines, whose limitations are discussed in the next section. Instead, we follow a `knowledge spatial' approach, using text embeddings which reflect the semantic relationships of the research literature.

An important distinction to make is that we are seeking a measure of the overall `breadth', or area, of covered research knowledge space of a researcher, rather than a measure of frequency of switching between different topics (cf. \citeproc{ref-zeng2019increasing}{Zeng et al., 2019}). This is because previous studies have shown that researchers can be continuously and concurrently active in multiple topics (\citeproc{ref-glaser2015bibliometric}{Gläser \& Laudel, 2015}; \citeproc{ref-hellsten2007self}{Hellsten, Lambiotte, Scharnhorst, \& Ausloos, 2007}; \citeproc{ref-cetina1999epistemic}{Knorr Cetina, 1999}; \citeproc{ref-simonton2004creativity}{Simonton, 2004, p. 79}). Thus, there need not be any `movement' from one area to another in the sense of having to discontinue work in one topic to take up work in another one.

We investigate the external criterion validity of potential measures of epistemic breadth primarily by comparing scientists of known high epistemic breadth to matched scientists of unknown (random) epistemic breadth who are otherwise similar. A useful measure of epistemic breadth should separate these two groups to some extent. We chose to compare these two groups due to pragmatic reasons, as we were not able to identify a group of scientists known \emph{a priori} to be highly specialized, that is, have low epistemic breadth.

The contribution of this study is therefore primarily methodological. We review proposed indicators for quantifying the extent of researchers' knowledge portfolio and identify several weaknesses of these methods. To overcome them, we introduce a novel knowledge space approach to the measurement of epistemic breadth. The main contribution is the validation of this proposed method. Once a first validated method is available, reliable empirical research on the distributions of epistemic breadth of various categories of researchers can be carried out, such as possible differences between disciplines, between career stages, differences across types of research institutions and sectors. Rigorous studies of the relationships between researchers' epistemic breadth and other social and cognitive properties and traits also become possible.

We continue this paper with a review of earlier work on building and using indicators for quantifying the expanse of researchers' topic portfolios, which leads to the motivation for introducing our own approach. Following this, we describe a novel knowledge space approach to measuring epistemic breadth, its current implementation with several considered variations, and the independent external data we use for validation of the proposed approach. Next, we report the results of our validation exercise and finally discuss these findings.

\section{Literature review}\label{literature-review}

There is a growing literature on the phenomenon of researchers expanding their area of study, or conversely, specializing, and related concepts. This literature uses a variety of different terms for these phenomena and competing measurement approaches have been introduced but no consolidation in terminology and methodology has been achieved yet. Many studies introduce a new measure for their conceptualization of the phenomenon without establishing its validity. We selectively review a number of representative studies, without claiming exhaustiveness, to illustrate the current state of research. We were able to identify two major directions in indicator construction, a classification system approach and a publication clustering approach, and we structure our review accordingly. A further approach is surveying researchers directly about their own perceptions of epistemic breadth (e.g. \citeproc{ref-hargens1986migration}{Hargens, 1986}). While this would generate valid and reliable data, it is rarely attempted, likely because of the costs and the necessarily limited comprehensiveness due to sampling.

\subsection{Work using classication systems}\label{work-using-classication-systems}

The first stream of literature follows the paradigm of using a pre-existing scientific literature classification system, often a disciplinary one, and constructing indicators that exploit the assignments of classes to papers or journals.

Jia, Wang, \& Szymanski (\citeproc{ref-jia2017quantifying}{2017}) measure `research interest change' of physicists as the cosine similarity of Physics and Astronomy Classification Scheme (PACS) class vectors of papers from the beginning and end of careers. They find that physics researchers have a small number of core topics they engage with repeatedly and a larger number of peripheral topics of lower importance to them. The sequence of engagement with topics is not random but rather clustered in time. Recently studied topics are more likely to be revisited than earlier ones. The particular subjects which researchers change to are typically relatively closely related to their previous subjects rather than epistemically distant ones. Hence, differences of the subjects are discussed on the basis of the temporal aspect, that is, as to whether they are novel or related to existing research, rather than how these topics are different from each other. It is possible that this overlooks situations in which a researcher engages with divergent topics concurrently.

Tripodi et al. (\citeproc{ref-tripodi2020knowledge}{2020}) investigate how physicists diversify their research topic portfolio, for which they use APS publication and PACS classification data. Influential factors for researchers to expand into new fields are how closely related the knowledge is and how many researchers who are already active in the new field they can collaborate with. The social relatedness factor, derived from the co-authorship network, is the dominant one of the two. In other words, within physics, the constraining factor for being able to engage with new topics is not how related a new topic is to a researcher's prior topics, but how well the researcher can access potential collaborators already familiar with the new topic.

Enduri, Reddy, \& Jolad (\citeproc{ref-enduri2015does}{2015}) also use PACS codes to measure topic diversity of physics papers and authors, proposing a topic diversity indicator which takes advantage of the hierarchical tree structure of the classification system. To compute the class diversity of an author, the diversity of the union of all PACS classes of all of the author's publications are considered. Some increase in the fraction of high-diversity authors across the observation period is reported. Moderate diversity papers are cited more frequently than those with low or high diversity.

Abramo, D'Angelo, \& Di Costa (\citeproc{ref-abramo2017specialization}{2017}) investigate the degree to which researchers specialize or diversify their research activity and how such tendencies vary across disciplines on the scale of the whole science system. High diversity is taken to be synonymous with interdisciplinarity. Their operationalization uses the Web of Science Subject Categories of the journals in which researchers published. The study found that the extent of diversification of researchers depends largely on their main discipline, such that for example the diversification of mathematicians is low and that of chemists is high.

Chakraborty, Tammana, Ganguly, \& Mukherjee (\citeproc{ref-chakraborty2015understanding}{2015}) measure the topic diversity and variation of computer science researchers over time with entropy of the class assignments of their publications according to an algorithmic classification system. They report that researchers increase the number of topics they work on in a given year in the earlier part of their careers while this figure is stable at a high level in the later part, in agreement to results of Zeng et al. (\citeproc{ref-zeng2019increasing}{2019}). It is further noted that scientists can be categorized according to the pattern of their topic changes into four groups and that those who work in particularly diverse fields over their career, but concentrate on relatively few fields concurrently, have empirically higher citation impact than those of other diversification categories.

Lawson \& Soós (\citeproc{ref-lawson2014thematic}{2014}) re-use a global map of Web of Science journal Subject Categories in which the position of categories is determined by bibliographic coupling relationships of categories, so theirs is a somewhat modified classification-based method. The thematic profile of a researcher is projected onto this map depending on which journals the researcher published in and into which Subject Categories these journals are classified. Thus, the profile reflects the areas of research of a researcher, their importance in her work and their relations to each other in terms of content similarity. The authors reinterpret a numerical `multidisciplinarity' diversity measure as `thematic mobility'. In a regression analysis they find that thematic mobility of UK engineering academics increases with academic rank, but is not associated with job mobility.

\subsection{Work using publication clustering}\label{work-using-publication-clustering}

The second stream of literature makes use of publication-level clustering and the application of indicators to the structure of the clustering solution, where clusters are interpreted as topics.

Zeng et al. (\citeproc{ref-zeng2019increasing}{2019}) propose to measure statistics of `topic switch' of physicists by clustering of the bibliographic coupling network of individual authors' publications. They find statistically significant non-random community structure in these networks such that the analyzed productive authors typically have a small number (3 or 4) of major topics important for them and a larger number of small topics or thematically isolated papers they work on infrequently. This is a historically stable pattern. The number of major topics that scientists concurrently work on varies with time: it is low in the early career, peaks around the 20th year of their career, then slowly decreases. This average number of concurrent topics of researchers has increased throughout the last century. The productivity-normalized probability of switching between those topics is low and increasing in the early part of a career but high and stable in the later part.

Franzoni \& Rossi-Lamastra (\citeproc{ref-franzoni2017academic}{2017}) cluster physicists' articles based on title and abstract text similarity. According to this study, researchers continuously extend the thematic diversity of their research throughout their careers and obtaining tenure is associated with a particular increase in diversity. The authors theorize that the job security of tenure permits the pursuit of a more risky strategy of diversification of research themes.

Hellsten et al. (\citeproc{ref-hellsten2007self}{2007}) consider the phenomenon of researchers' `field mobility' and test if such mobility can be detected through author self-citation analysis. In their case study, they focus on the largest connected component of the self-citation network (ignoring singletons and smaller components) and selectively remove a number of nodes to minimize the connectivity of the sub-network. For the three identified subfields of activity of an example researcher they observe clear differences in keywords and co-authors between the publication sets.

\subsection{Critical assessment}\label{critical-assessment}

The reviewed literature on measures of epistemic change and research profile expansion have mainly used two approaches: publication clustering and classification systems, neither of which is without problems.

In the first approach, the assignments of publications to a science classification system are taken as basic data and the distribution of publications over classes --and sometimes the structure of the classification system-- is used to derive measures of epistemic breadth. For example, Enduri et al. (\citeproc{ref-enduri2015does}{2015}), Tripodi et al. (\citeproc{ref-tripodi2020knowledge}{2020}), Jia et al. (\citeproc{ref-jia2017quantifying}{2017}), Yu, Szymanski, \& Jia (\citeproc{ref-yu2021become}{2021}) have used the Physics and Astronomy Classification Scheme, Pramanik, Gora, Sundaram, Ganguly, \& Mitra (\citeproc{ref-pramanik2019migration}{2019}) used Microsoft Academic's system, the Mathematical Subject Classification was used by Basu \& Wagner Dobler (\citeproc{ref-basu2012cognitive}{2012}), and Web of Science Subject Categories (WoS SC) by Lawson \& Soós (\citeproc{ref-lawson2014thematic}{2014}) and Abramo et al. (\citeproc{ref-abramo2017specialization}{2017}). Some classification systems are discipline-specific which strictly limits their applicability. Using a discipline-specific classification system, even if just for studying epistemic breadth in one discipline, comes with the unstated and implausible assumption that researchers never study topics beyond the discipline covered by the classification system. Values of epistemic breadth for such cases are inevitably underestimated in these studies, undermining their validity. Studies using universal classification systems avoid this issue but expose themselves to other threats to validity. Subject classes can have differing scopes, some being very narrow, others broad (between-class heterogeneity). If a researcher has published in one predefined class, this does not mean that their research activity or interest covers the whole of the cognitive content of that class. Publications classified with the same class or set of classes appear equivalent in this approach even though they may actually address quite distinct issues within a class (within-class heterogeneity). This is particularly a problem in broader-scope categories and journal-level systems such as the WoS SCs.

In the second approach, individual scientists' publications are algorithmically clustered into discrete topic groups as a basis for calculating topic changes (\citeproc{ref-franzoni2017academic}{Franzoni \& Rossi-Lamastra, 2017}; \citeproc{ref-hellsten2007self}{Hellsten et al., 2007}; \citeproc{ref-zeng2019increasing}{Zeng et al., 2019}). These have been \emph{ad hoc} in their choice of clustering algorithm and setting of parameters. Different clustering results can be obtained by choosing between the many available clustering algorithms, by the deliberate setting of parameter values, and through the influence of many other methodological factors (\citeproc{ref-struck2025dependencies}{Struck, 2025}). It is usually not clear how stable the studies' findings are to this degree of freedom in investigator methodological choices. A notable exception is the study of Zeng et al. (\citeproc{ref-zeng2019increasing}{2019}), which used a variety of methodological variations to ensure their results are robust and not artifacts of their choice of methods.

In both approaches, two researchers may change fields equally often, but if the epistemic distance between fields, observed as classification system or cluster solution categories, of one of them is small while the distance between fields of the other is large, this meaningful difference would remain unobserved if epistemic distance between categories remains unspecified or is not incorporated into a measure. This is partially addressed with measures that exploit the hierarchical tree structure of classification systems, such as the use of Weitzman diversity for PACS in Enduri et al. (\citeproc{ref-enduri2015does}{2015}). Another approach is to explicitly measure category distance or publication distance and incorporate this measurement into the calculations (e.g. \citeproc{ref-glaser2015epistemic}{Gläser, Heinz, \& Havemann, 2015}; \citeproc{ref-lawson2014thematic}{Lawson \& Soós, 2014}; \citeproc{ref-tripodi2020knowledge}{Tripodi et al., 2020}; \citeproc{ref-yan2015understanding}{Yan \& Lagoze, 2015}).

For the purpose of calculating epistemic breadth, in both approaches the calculation of topic changes is moved from the level of publications to that of categories, a higher and less granular level of abstraction. These categories are not identical with the content of publications but represent larger conceptual constructs, which may be a poor or a good match for a given publication.

The reviewed studies have proposed measures of constructs identical or closely resembling our conception of epistemic breadth, using a variety of different terms such as `cognitive mobility', `research topic switching', `research interest change/evolution', `change in research agenda/direction', and `research portfolio diversification'. Terminology is evidently unorganized. There is also a notable lack of cohesion and knowledge accumulation in this area of research because the measures proposed by one research group are not used further, are not independently verified, revised or replaced by other groups. Those active in this area have been working in methodological isolation, preferring to use their own measures without showing why their particular method is preferable to others. This state of affairs makes the various results difficult to compare and align. Even more so, as these studies refer to different (though apparently similar) concepts.

Finally, none of these measures has been sufficiently validated. The reviewed studies jump from indicator design to application, skipping validation. The indicators which were introduced to measure this phenomenon have not been convincingly demonstrated to really measure what they are claimed to measure by means of independent empirical data. This is not to say that extant methods are not theoretically sound or that the studies' results are necessarily flawed. But without empirical validation of measurement methods, it is questionable which results are really reliable (\citeproc{ref-gingras2014criteria}{Gingras, 2014}; \citeproc{ref-markus2010construct}{Markus \& Lin, 2010}).

Indicator validity is not just nice to have, it is an indispensable precondition for useful empirical research. To understand the importance of the point, consider the methodological situation in a thematic area closely related to the present one. There is a vast literature of bibliometric studies on the construct of interdisciplinarity which has had considerable policy impact. Reviewing a number of different measurement methods and comparing their result values, Wang \& Schneider (\citeproc{ref-wang2020consistency}{2020}) arrive at the following sweeping conclusion:

\begin{quote}
The validity and robustness of interdisciplinarity studies using bibliometric methods should be questioned. As it is, measures and their values are inconsistent and non-robust. This can lead to an untenable situation where the choice of (arbitrary) measures determines the degree of interdisciplinarity, but not the underlying nature of research which they are supposed to characterize. We therefore suggest that future studies on interdisciplinarity focus more upon the theoretical and measurement frameworks, and put more effort into examining the validity and relations between the definition and the use of measures. In addition, we recommend that we simply stop using the current interdisciplinarity measures in policy studies, as they have no warrant.
\end{quote}

From our critical assessment of the literature on the phenomenon of the structure and dynamics of researchers' topic profiles, we conclude that a similar state of affairs prevails in this area.

In summary, the issues left insufficiently addressed by proposed bibliometric measures of epistemic breadth that motivate our new methodological contribution are:

\begin{itemize}
\tightlist
\item
  various problems of category-publication mismatch in classification and clustering systems
\item
  unacknowledged arbitrariness and flexibility in clustering algorithm and parameter setting choices
\item
  applicability to all disciplines
\item
  lack of empirical validation.
\end{itemize}

\section{Methods and data}\label{methods-and-data}

\subsection{Approach}\label{approach}

Similar to Gläser et al. (\citeproc{ref-glaser2015epistemic}{2015}), we propose to forgo to try to explicitly identify discrete categorical topics for publications to subsequently use these topics (or classes/clusters) as an indirect way to operationalize epistemic breadth. While Gläser and colleagues studied hypothesized decreases in cognitive diversity of research fields, not the structure of individuals' research portfolios, we concur with their argument that assigning publications to discrete topics is at odds with the properties of scientific topics. Instead we use a `knowledge spatial' approach. In this abstraction, each publication has its own precise, unique, and deterministic location in a common knowledge space depending on its content and semantically meaningful distances between publications based on their content can be calculated.

The concept of a knowledge space refers to an abstract mathematical representation of a body of knowledge in which units of knowledge can be located and whose properties reflect the cognitive structure of the represented knowledge, forming a computational model of a body of knowledge. This is primarily realized by the construction of the knowledge space in such a manner that measured distances between representations of knowledge units in knowledge space, namely coordinates of publications or texts, accurately reflect the semantic similarity of their respective subject matters.

The principal insight motivating our approach is as follows. A researcher whose publications are all located close to each other in knowledge space has low epistemic breadth -- she or he is highly specialized, all publications are within the same topic. The publications of a researcher who worked on many divergent topics will be spread out across a wider `area' in knowledge space and publications will have overall lower similarities (greater distances) to one another.

\subsection{Publication data}\label{publication-data}

We use bibliometric data from Scopus, a large multidisciplinary scholarly publication and citation database (\citeproc{ref-baas2020scopus}{Baas, Schotten, Plume, Côté, \& Karimi, 2020}). As we study individual researchers, it is crucial to have author-disambiguated publication data of high quality. Scopus' algorithmically disambiguated and partly manually corrected author data (\citeproc{ref-baas2020scopus}{Baas et al., 2020, p. 379}) has been demonstrated to be of sufficient quality for large-scale studies (\citeproc{ref-aman2018does}{Aman, 2018}; \citeproc{ref-kawashima2015accuracy}{Kawashima \& Tomizawa, 2015}).

\subsection{\texorpdfstring{\texttt{SPECTER} semantic text embeddings}{SPECTER semantic text embeddings}}\label{specter-semantic-text-embeddings}

For the measurement of semantic similarity between documents we build on \texttt{SPECTER} scientific document embeddings (\citeproc{ref-cohan2020specter}{Cohan, Feldman, Beltagy, Downey, \& Weld, 2020}). The \texttt{SPECTER} model is a language model for scientific text, pre-trained on both the title and abstracts and the citation link relationships of a corpus of scientific research papers. The model is provided ready to use and requires some text of a document as input, for which we use title and abstract. It does not need citation data at this stage, so the returned vector representation is stable and does not change if the citation environment of a document changes.

The model output for a document is a dense real-valued vector representation of a fixed length which encodes the textual meaning of the input text. The vector representation can be seen as the coordinates of the location of a text in a high-dimensional abstract vector space learned during training -- what we call here `knowledge space'. The semantic similarity of two text embedding vectors from the same model can be obtained by calculating their cosine. While the dimensions do not carry interpretable meaning, embedding models are designed to learn a representation space in which semantically similar texts will have close locations even if they do not use the same lexical terms. \texttt{SPECTER} was chosen because it was extensively validated and benchmarked, is the current state-of-the-art method (\citeproc{ref-cohan2020specter}{Cohan et al., 2020}), and is stable for a given input -- meaning results are reproducible for individual documents across different corpora. However, if better semantic text embedding models are developed in the future, they could easily be used instead of \texttt{SPECTER} in our methodology if the improvements justify this change.

\subsection{Validation of candidate measures with external data}\label{validation-of-candidate-measures-with-external-data}

For the indicator validation, we use primarily the publication data of researchers funded through the Cross-Disciplinary Fellowships instrument of the Human Frontier Science Program (HFSP). The HFSP is run by the International Human Frontier Science Program Organization, an international organization established 1989, whose members include a number of countries and the EU. The mission of the HFSP is to fund highly ambitious basic research into frontier life science topics through several instruments, mainly project grants and researcher fellowships. We use data on their Cross-Disciplinary Fellowships program. The HFSP states that ``Cross-Disciplinary Fellowships (CDF) are for applicants who hold a doctoral degree from a non-biological discipline (e.g.~physics, chemistry, mathematics, engineering or computer sciences) and who have not worked in the life sciences before.''\footnote{\url{https://archive.is/Nknoq}}
Due to this selection criterion, this group of researchers is known to have at least one major event of research topic change of large extent, from any other different science field to biology, at a clearly defined time point. We collected the names of CDF recipients of the period 2005 to 2016 from the HSPF website and manually matched them to Scopus author IDs. For 94 of these funded researchers, we were able to identify the corresponding Scopus author profiles. For the comparative validation of the measure, we tried to match each CDF recipient with one randomly selected researcher of similar characteristics from the target population of Germany-affiliated scientists (explained below). The pairs were required to match on

\begin{itemize}
\tightlist
\item
  primary research area: the Scopus All Science Journal Classification category most often assigned to their publications
\item
  the number of publications ± 10 \%
\item
  the time period of actively publishing ± 1 year of first and last recorded publishing year, respectively.
\end{itemize}

We use the matching pair strategy to minimize the adverse influence of factors such as career length, main research area and productivity on measure validation. It was possible to find matches for 86 CDF recipients.

In the further course of the project we want to study researchers in the STEM research domains who have met a set of inclusion criteria chosen to ensure sufficient publication data at the individual level for meaningful statistical analysis. The researchers needed to have any affiliation in Germany in their publishing career and needed to have contributed to at least ten publications.

\subsection{Validation of measure with internal data}\label{validation-of-measure-with-internal-data}

We also use internal data, that is, bibliometric data on the author level, to further validate candidate measures of epistemic breadth. Hellsten et al. (\citeproc{ref-hellsten2007self}{2007}) have proposed to use author self-citations for studies of epistemic variety and change. The theoretical reasoning is that publications of an author that are connected by citations, compared to unconnected ones, are relatively more likely to be concerned with the same topic as the author or their working group successively builds on previous research in continuing work on a larger project. Conversely, publications on topics peripheral to an author's major interests, in which they have been only incidentally and marginally involved, are less likely to cite previous publications or to be cited by later publications of the author. It follows that, \emph{ceteris paribus}, highly specialized authors (low epistemic breadth) should have a densely self-citation connected publication set, while the publications of authors of high epistemic breadth should show fewer self-citation connections. We consider the self-citation network approach useful to validate our method but are less convinced about the suitability of this data as the primary data for measures of epistemic breadth because authors' self-citation propensities are likely to be influenced by idiosyncratic personality factors, thereby potentially making any quantitative measurement constructed from this data noisy. Consequently, we expect moderate, but not high, correlations between self-citation indicators and candidate knowledge space measures of epistemic breadth.

For each author in our larger analysis data set we have calculated the following three self-citation indicators. First, the self-reference rate is the fraction of the number of cited references of which the focal author is also an author (self-references) divided by the total number of cited references. Second, the realized self-reference rate is a refinement of the first indicator. In this variant, the quotient is calculated as the number of self-references divided by the number of previously published own papers of the author that could potentially have been self-cited. The third indicator is the normalized average size of the connected components of the author's self-citation network. For this, the number of connected components of the author self-citation network is calculated. The number of publications is then divided by the number of components. An author who never self-cited their own work would have an average value of 1 for this indicator -- each component is a singleton. An author who always self-cites all previous works or the one immediately preceding publication of any new publication, such that all publications are connected, would have a value equal to the number of publications. This value is then normalized by dividing it by the number of publications to control for size effects. Otherwise authors with larger publication volumes would tend to have higher values. By normalization all values are constrained to the interval from zero to one. Unlike Hellsten et al. (\citeproc{ref-hellsten2007self}{2007}), who restrict the analyzed data to the largest self-citation connected component of the network, we consider all publications of an author. We report correlations between the self-citation indicators and the preferred candidate measure of epistemic breadth obtained from the validation with external data. We give priority to the external data because being a CDF Fellowship recipient is more directly and concretely associated with epistemic breadth.

\subsection{Candidate measures}\label{candidate-measures}

As the natural starting point and baseline for the development of epistemic breadth measures we use the simplest possible measure in the knowledge space framework, the arithmetic average of the cosine similarities between all unique pairs of papers of an author. This candidate measure uses the full information of the similarity network in the most straightforward way. Any other more sophisticated measure would require additional theoretical justification and should empirically show better results in order to be preferred.

The second considered candidate measure works by first calculating an intermediate score for each publication and then averaging those. Since the goal is to quantify the analog to an `area' in knowledge space covered by an author's publications, we test the minimum similarity of a publication to any other publication, that is, the furthest neighbor similarity, as an intermediate score. In contrast, if we take as an intermediate publication score the maximum similarity of a publication to any other (nearest neighbor similarity) and then average these, that measure should show worse discrimination, as it only takes into account the very local neighborhood of each publication. If an author publishes several papers in various divergent topics, only the papers within a topic, being similar to each other, would appear as nearest neighbors. But the similarity values of papers in different topics would never enter the calculation, while it is these which should best characterize the extent of covered knowledge space, as their distance can be interpreted as approximating the diameter of the researcher's knowledge contribution profile in the knowledge space. Therefore, we include this measure as a deliberate contrast to verify our assumptions, but not as a likely viable candidate measure for epistemic breadth.

We also consider a network theoretic measure, the average shortest path length of the publications' \texttt{SPECTER} similarity network. In the similarity network there is a direct path between any two nodes. But because of the different distances, here operationalized as (1 -- similarity), the direct path need not be the shortest path.

Finally, we also study a robustness modification to the furthest neighbor candidate measure. Due to the design of that measure, a single paper which is topically very distant from all other papers can appear as the furthest neighbor to all other papers even if the focal author's involvement in the paper was very minor. Authors usually contribute unequally to co-authored papers and papers to which authors have contributed only very little are less important to their body of work than those of which they were main contributors (cf. \citeproc{ref-glaser2015bibliometric}{Gläser \& Laudel, 2015, p. 311}). It is further probable that authors contribute only slightly to papers which are peripheral to their primary research interests and can contribute much more to papers in their core topics of interests. This presumed correlation between the degree of contribution or involvement in a co-authored study and the proximity of the study's topic to the researcher's core interests could lead to overemphasis of the influence of topical outlier papers with small author contributions. Hence, we modify the minimum per-paper similarity measure by calculating the weighted average of intermediate scores instead of the simple average. The weights are determined from author position and author count by harmonic author credit scoring (\citeproc{ref-hagen2008harmonic}{Hagen, 2008}) which has been shown to statistically approximate the relative contributions in co-authored papers in the sciences (\citeproc{ref-donner2020validation}{Donner, 2020}). This method assigns the greatest contribution share for a paper to the first author and ever smaller shares to the following co-authors in the author list. All shares sum to unity which also means that co-authors at the same absolute position of a paper with less co-authors receive more credit than those on a paper with more co-authors.

To statistically quantify how well the treatment and control groups can be distinguished by the candidate measure we use Cohen's \emph{d} statistic, a standard effect size metric which expresses the difference of group means in standard deviation units.

\section{Results}\label{results}

\subsection{Qualitative assessment of the soundness of the knowledge space approach by visualizations of key knowledge space sections}\label{qualitative-assessment-of-the-soundness-of-the-knowledge-space-approach-by-visualizations-of-key-knowledge-space-sections}

We want to initially provide qualitative, visual evidence that the knowledge space approach of using the semantic embedding similarity network of authors' publications to assess their epistemic breadth is fundamentally sound. To this end, we provide visualizations of some of the data from our matched author pairs. We created visualizations of the local knowledge space structure of each author pair's publications with Multidimensional Scaling (MDS), a standard dimension reduction method for visualizing high-dimensional data in 2 or 3 dimensions. In our 2-D plots below, each paper is shown as a letter symbol in a common 2-dimensional space, optimized by the MDS algorithm to distort as little as possible the overall similarity relations that enter each plot. This means that the positions of symbols representing papers in these plots are determined only by the similarity relations of the papers to each other in the common global \texttt{SPECTER} knowledge space. Papers that are very closely semantically related will be positioned next to each other. The two dimensions have no intrinsic meaning associated with the semantic content of paper embeddings, they are purely algorithmic constructions of the MDS procedure. Moreover, as only the embeddings similarity data of the papers of the two authors of one matched pair is used to generate each visualization, they show local, incomplete parts of the global common knowledge space and these parts are different for each author pair.

When visualizing two matched authors' publications together in one plot in this way, we would expect the following, if the fundamental approach is sound. The publication symbol points of the two authors should usually form two point clouds that are at least partly distinguishable, as the paired authors would have worked on different specific topics, hence the visualized surrogates of their papers should occupy distinct regions and not be totally intermixed. However, some partial overlap of point clouds of the two authors' publications can also be foreseen, as they were matched also on their primary discipline and may have coincidentally worked on closely related problems. Secondly, and more decisive for the qualitative appreciation of our proposed operationalization, the point clouds of the CDF funded researchers, who have been funded to switch disciplines, should be more dispersed than those of the matched control researchers.

\begin{figure}
\centering
\includegraphics[width=1\linewidth,height=\textheight,keepaspectratio]{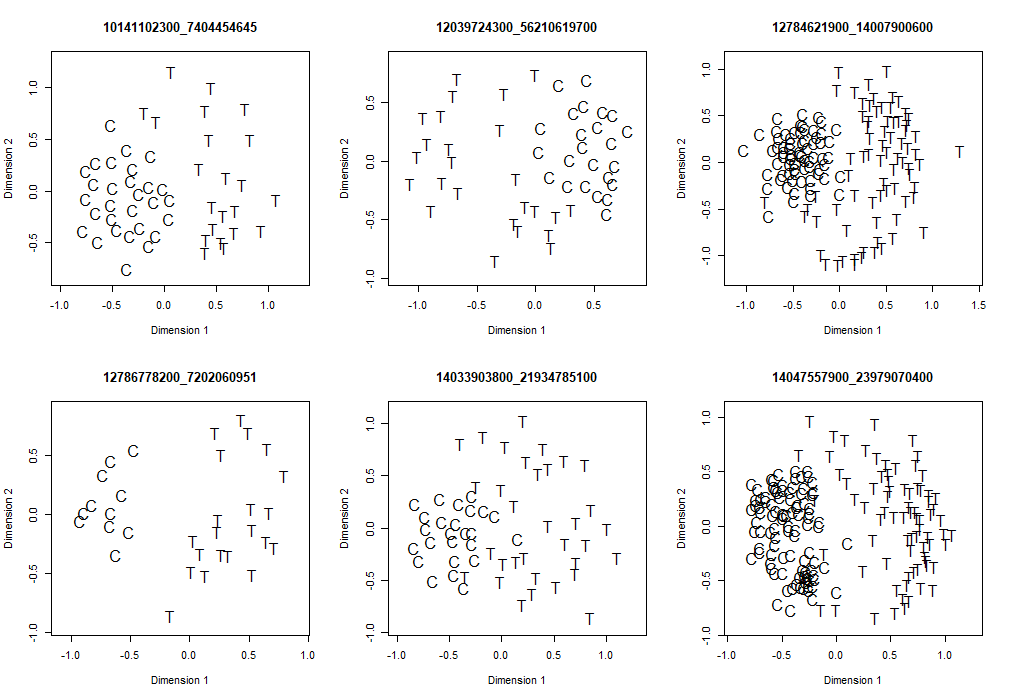}
\caption{Example MDS visualization of six matched treatment-control author pairs' paper positions in local knowledge space. Labels: T: treatment researcher publications (changed discipline). C: Matched control researcher publications.}\label{fig:fig1}
\end{figure}

As a first illustration of the approach, Fig. \ref{fig:fig1} shows MDS visualizations of the \texttt{SPECTER} cosine similarities between all papers of matched treatment and control author pairs for six randomly chosen pairs. The plot panel headings are the Scopus author IDs of the two matched researchers. Here we only show six pairs so that the plotting panel for each pair can be large enough for easy inspection. A visualization of the full data set is provided below. It can be seen that the publications of CDF funded researchers (labeled ``T'') generally form less dense point clouds than those of the comparable, matched control researchers (labeled ``C''). The publications of each researcher also mostly cluster relatively closely together in one half of the visualized region of knowledge space. This lends initial qualitative support to our approach.

\begin{figure}
\centering
\pandocbounded{\includegraphics[keepaspectratio]{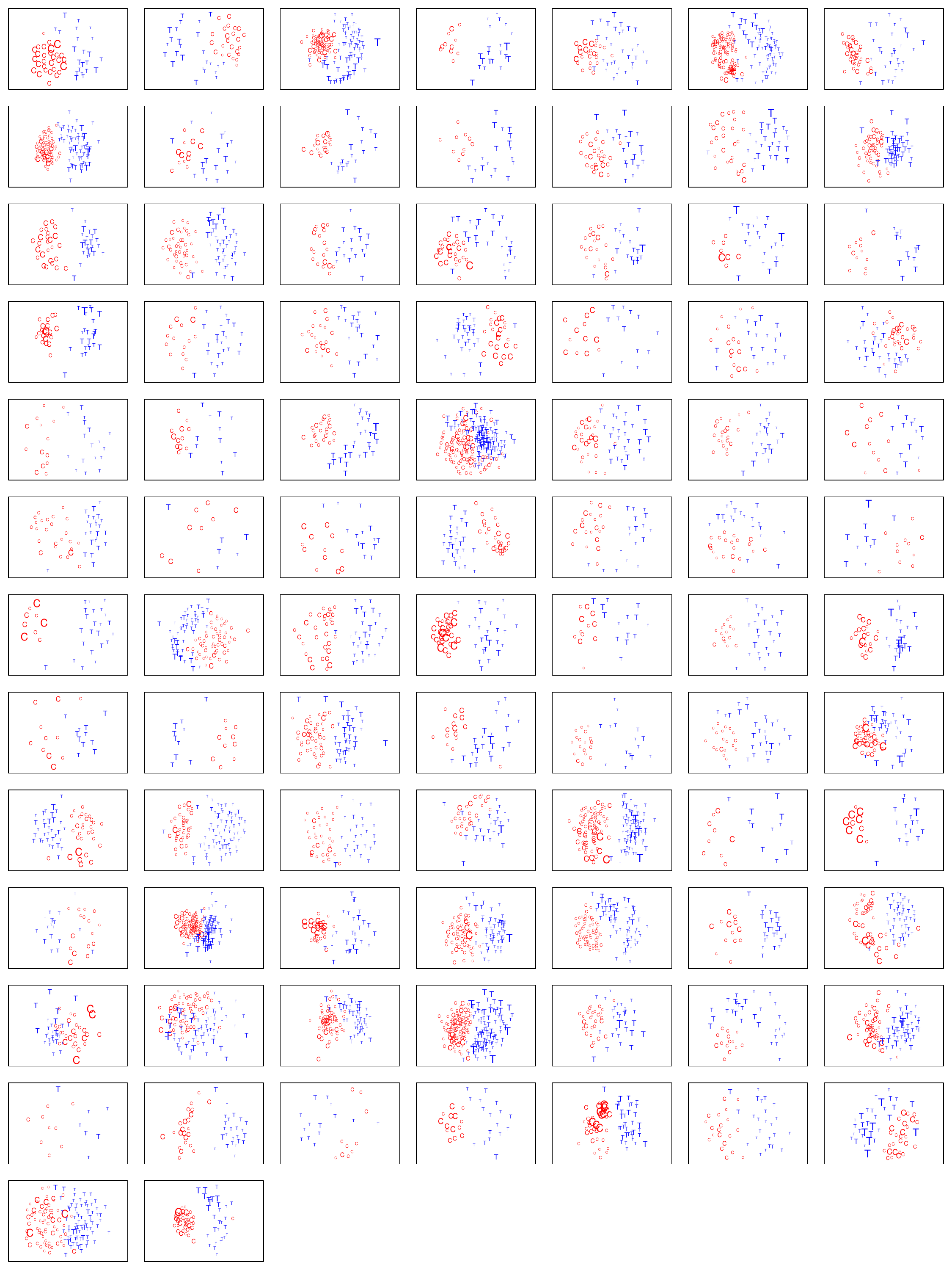}}
\caption{\label{fig:fig3}MDS visualization of all treatment-control author pairs' paper positions in local knowledge space. Labels: T: treatment researcher publications (changed discipline). C: Matched control researcher publications.}
\end{figure}

A visualization of the complete validation data set of all 86 author pairs using the same technique is shown in Fig. \ref{fig:fig3}. In this figure, since the panels are smaller, the treatment and control paper position symbols have been given different colors and the symbols have also been scaled by approximate author contribution as inferred from author position and co-author count of papers. Symbol size is proportional to harmonic co-author credit scores (\citeproc{ref-hagen2008harmonic}{Hagen, 2008}; \citeproc{ref-hodge1981publication}{Hodge \& Greenberg, 1981}), that is, to the weighting factors we have chosen to test for the above discussed robustness modification of the furthest neighbor candidate measure for epistemic breadth. One can see different variations on the expected patterns. In some cases, the two point clouds are fully separated, in other cases they show much overlap. Again, the plotted symbols for the papers of discipline-switching (CDF funded) researchers are usually more widely dispersed than those of the matched controls. One can also discern a tendency for paper symbols in the centers of the point clouds to have been assigned larger symbols, that is, having presumably greater relative involvement of the focus author among their co-authors, such as being first author rather than later author and having smaller author teams, rather than bigger ones. In some cases, the CDF funded researcher's point clouds are even quite clearly divided into two or more unconnected regions, indicating semantically distant topics.

\subsection{Average similarities of discipline-switching and control authors' papers}\label{average-similarities-of-discipline-switching-and-control-authors-papers}

Next, we provide a high-level assessment of the semantic paper similarity profiles of discipline-switching and control researchers. We have calculated the average cosine similarity of the \texttt{SPECTER} embeddings of the papers of each author. That means, for each paper of one author, we calculate the similarities to all other papers, and average these values. What we would expect is, first, that the overall distributions of average similarities of CDF funded (discipline-switching) researchers and control researcher are different, specifically that the controls have higher average similarities. This is because they should, as a group, have lower average epistemic breadth, so their papers should be more similar to each other, relative to CDF fellowship recipients. Second, also within each pair of authors, the controls should have higher average \texttt{SPECTER} embedding similarity compared to their specific matched CDF fellow.

\begin{figure}
\centering
\pandocbounded{\includegraphics[keepaspectratio]{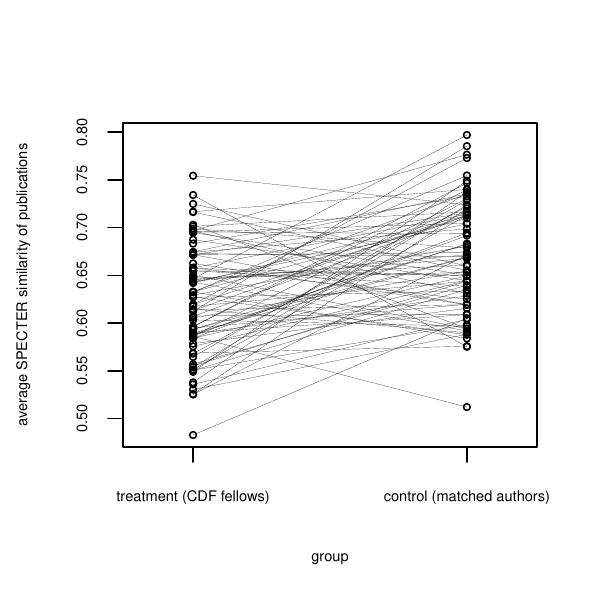}}
\caption{\label{fig:group-avg-comp}Average cosine similarity of publications of CDF and matched control authors. Lines indicate matched pairs.}
\end{figure}

Fig. \ref{fig:group-avg-comp} shows the author average cosine similarities of papers of matched treatment-control pairs. The distribution of average similarities for treatment authors (discipline-switchers) is on the left of the plot, that of the control research is on the right. A line has been drawn between the points for the averages of each matched pair. The similarities between papers are larger on average for the control group (mean: 0.67, SD: 0.06) as a whole compared to the treatment group (mean: 0.63, SD: 0.06). For the majority of pairs (72 \%), the value for the control group is also greater than that of the treatment group, visualized by a connection line that is angled upward from left to right. This provides a further confirmation that the text embeddings similarity approach in the knowledge space framework works as expected and, incidentally, that the most simple baseline candidate measure is able to measure epistemic breadth to some degree. However, some matched control authors exhibit lower average similarities than CDF fellows and in a fraction of pairs it is the control that has the lower similarity.

\subsection{External data validation: discipline-switching CDF fellow}\label{external-data-validation-discipline-switching-cdf-fellow}

Having thus established some support for the soundness of our approach, we can move on to the main validation findings. Table \ref{tab:main-results} shows the main results of our study, the effect sizes (Cohen's \emph{d}) for the difference in candidate epistemic breadth measures. Cohen's \emph{d} is a standardized effect size measure, it measures the difference in group means in units of standard deviations. For this study, the interpretation of this effects size for a given tested candidate measure is how well a measure can discriminate highly epistemically broad authors from random authors (matched to various attributes but with \emph{a priori} unknown, random epistemic breadth).
Higher absolute values indicate better discriminating measures.

The results for the simple arithmetic average and similarity graph distance measure are very similar, in effect size and also the measured values of researchers, likely because in nearly all cases the shortest path is the same as the direct path and all papers in our paper similarity graphs are connected by edges. The absolute values of around 0.8 are considered large effect sizes.

The unweighted and author-contribution weighted furthest neighbor candidate measures show slightly higher (negative) effect sizes of around --0.85. Since we have relatively large confidence intervals, these small differences are not statistically significant. They tentatively support of theoretical reasoning for introducing the furthest neighbor modification. But they provide no support for the consideration of using approximated author contribution as a weighting factor.

Confirming our assumptions about nearest neighbors in the similarity graph not providing the relevant information, the nearest neighbor candidate measure's effect size is much smaller at about --0.55, which is still a medium effect size. None of our candidate measures completely separates the scores of treatment and control group. This is not concerning as the control group researchers were chosen randomly with respect to their epistemic breadth, and therefore include the full range of epistemic breadth values from lowest to highest, while the treatment group was chosen for its known elevated epistemic breadth, which does not mean they need to have the highest possible epistemic breadth.

To provide some context for interpreting the result values, a \emph{d} value of --0.85 can be converted to a correlation coefficient of 0.39. Absolute \emph{d} values in the .80s are generally considered large or very large.

On the basis of these results we conclude that the weighted furthest neighbor measure is an adequate measure of epistemic breadth to continue studying and validating. However, its effect size value is not clearly statistically significantly greater than those of the other candidate measures and our preference is partially informed by the theoretically higher robustness to topical outlier papers with minor author contributions.

\begin{table}

\caption{\label{tab:main-results}Comparison of performance of different candidate measures of epistemic breadth in distinguishing discipline-switching researchers from matched control researchers.}
\centering
\begin{tabular}[t]{ll}
\toprule
candidate measure & effect size (Cohen's d [95\% CI])\\
\midrule
arithmetic average (baseline) & -0.81 [-1.12, -0.5]\\
furthest neighbor per paper average & -0.87 [-1.19, -0.56]\\
weighted furthest neighbor per paper average & -0.85 [-1.16, -0.53]\\
nearest neighbor per paper average (contrast) & -0.54 [-0.85, -0.24]\\
similarity graph distance average & 0.82 [0.5, 1.13]\\
\bottomrule
\end{tabular}
\end{table}

\subsection{Internal data validation: correlations with self-citation indicators}\label{internal-data-validation-correlations-with-self-citation-indicators}

We now consider the correlations of the author-contribution weighted furthest neighbor candidate measure, WFN for short, with the three self-citation indicators for our larger analytical sample of authors who have had any German institutional affiliation in their careers. As the coverage of our Scopus data begins in 1996 we have excluded authors who have published in 1996 or 1997 in an attempt to include mostly authors whose publication records are not left-truncated in time. The data set comprises 179,298 cases. The Pearson correlation of the WFN measure of epistemic breadth to the simple self-references rate is \emph{r}~=~--0.04 (\emph{p}~\textless~0.01; 95\% CI: {[}--0.05, --0.04{]}). As higher values of WFN indicate more specialization, or less thematically broad knowledge profiles, this association suggests that more epistemically broad authors reference their own publications more, which is contrary to our expectation. Deferring an interpretation of this finding and moving on to the more sophisticated indicators, the correlation of WFN and the realized self-reference rate is \emph{r}~=~0.40 (\emph{p} \textless~0.01; 95\% CI: {[}0.40,~0.41{]}). Thus, once self-references are put into relation only to the number of actually existing own publications at any given time, which therefore could have possibly been self-cited, the picture looks very different. This value confirms that more specialized researchers with a relatively narrow thematic profile tend to reference their own prior papers more. This is in line with our expectation, as more specialized researchers find it easier to relate any given new publication to a greater number of earlier papers. The correlation between WFN and the normalized average size of connected components of researchers' self-citation network is \emph{r}~=~0.40 (\emph{p} \textless~0.01; 95\% CI: {[}0.39,~0.40{]}). This alternative indicator corroborates the preceding result. Researchers who are more epistemically broad (lower WFN value) have self-citation networks of papers that are less densely connected and consist of more, and smaller, components. In light of these additional results it seems likely that the unexpected small negative correlation for the simple self-reference rate is an artifact, possibly due to a size effect.

\section{Discussion}\label{discussion}

Epistemic breadth of researchers --the extent to which they study multiple, possibly divergent topics throughout their career-- is an important property of their research agendas. Growing research interest in this area has produced a literature on equivalent or closely related concepts which our critical review has characterized as fairly uncoordinated, resulting in rather different ways of measuring this property. None of the various proposed measures for the concept has been validated with empirical data as of yet. This means that is unknown if these indicators actually measure what they purport to measure, undermining the solidity of their findings and the cumulativeness of this collective research endeavor. A number of proposed methods focus on topical diversity or switches within a single discipline only. Others do not take into account the semantic distance between topics, however operationalized. Most methods use discrete categories for topics, an approach which has considerable drawbacks.

We have proposed a novel knowledge space approach to measuring epistemic breadth with bibliometric data to address these shortcomings. In particular, our method is based on the semantic similarity of researchers' publications, namely the epistemic distances between publications in one common knowledge space. Prior research has not used the micro-level cognitive-structural properties of individual publications but relied on higher order aggregations such as clusters or classification categories, which instantiate properties that member publications have in common, rather than what distinguishes them individually. This level of indirection puts doubt on how accurately researchers' knowledge domains can possibly be measured in such approaches.

Using empirical data on researchers specifically funded for making a major change to their research topic portfolio and self-citation data, we have compared candidate measures within the knowledge space approach as to how well they capture the concept of epistemic breadth. Our main validation exercise was done with publication data of of researchers who were specifically funded to change their discipline of research to biology from any other science discipline. We found a strong effect size for proposed indicators of epistemic breadth when comparing the epistemic breadth values of this treatment group of discipline-switching researchers with a group of matched researchers of random, unknown epistemic breadth. We were able to show that even a simple arithmetic average of papers' semantic similarities is a useful indicator of epistemic breadth. Better performance is achieved by a slightly more sophisticated indicator using per-paper furthest neighbor similarities.

Additional validation by comparison of self-referencing statistics of a much larger set of researchers and their epistemic breadth values for our preferred indicator revealed robust and substantial associations in the theorized directions, further substantiating the validity of this method.

However, we do not claim that this is the final word in indicator development for epistemic breadth -- hopefully it is a significant step in the right direction. We need to acknowledge several limitations of our study, some of which may hopefully be remedied by follow-up research. We have used the best external validation data that we were able to find. Yet, our data only includes highly epistemically broad junior researchers who switched from other science fields to biology. We are thus missing more precise criterion data along the whole range of epistemic breadth -- we had no researchers of known moderate or low epistemic breadth in our sample. Nor does it include senior researchers and scholars from the social sciences and humanities. This shortcoming could perhaps be addressed by collecting survey self-rating data from researchers. Next, we have only used a single semantic embeddings method to locate papers in common knowledge space. While we used a method that was state-of-the-art when we started our study, development in this area is fast and newer methods may be better. Our approach is in principle flexible, in that it is not tied to the embeddings method we used. As better methods become available, they can just be substituted, and the resulting modified measure needs then of course to be validated again. Finally, we have studied only a limited number of algorithmic variants for different candidate measures, other researchers may be able to come up with improved implementations.

We have not studied the methodological validity of the various similar proposed measurement methods that we have critically reviewed. Thus, we would like to encourage other researchers to further validate their approaches with similar or better validation strategies. That would make it possible to decide, based on effect sizes for different methods on the same or similar data sets, to decide which methods to use and with what confidence.

\section{Conclusion}\label{conclusion}

We have introduced \emph{epistemic breadth} as a key property for the description of researchers' research strategies, as it models the dimension of the thematic variation, from highly specialized to highly diversified. No method for the bibliometric measurement of the concept of epistemic breadth has so far been subjected to a validation study.

This study presents the first empirically validated bibliometric indicator for measuring epistemic breadth of the publication profiles of individual researchers. We showed that an approach which models publications as points in an abstract global knowledge space and utilizes the mutual semantic similarities of publications, that is, their distances in this knowledge space, as a foundation, can support the development of valid indicators of epistemic breadth.

We believe that formal empirical indicator validation is a crucial step for achieving cumulative progress in this and other topics of science studies.

\emph{Funding information}\\
This research was funded by German Federal Ministry of Education and Research, project M536400.

\section*{References}\label{references}
\addcontentsline{toc}{section}{References}

\phantomsection\label{refs}


\begin{CSLReferences}{1}{0}
\bibitem[\citeproctext]{ref-abramo2017specialization}
Abramo, G., D'Angelo, C. A., \& Di Costa, F. (2017). Specialization versus diversification in research activities: The extent, intensity and relatedness of field diversification by individual scientists. \emph{Scientometrics}, \emph{112}(3), 1403--1418. \url{https://doi.org/10.1007/s11192-017-2426-7}

\bibitem[\citeproctext]{ref-aman2018does}
Aman, V. (2018). {Does the Scopus author ID suffice to track scientific international mobility? A case study based on Leibniz laureates}. \emph{Scientometrics}, \emph{117}(2), 705--720. \url{https://doi.org/10.1007/s11192-018-2895-3}

\bibitem[\citeproctext]{ref-baas2020scopus}
Baas, J., Schotten, M., Plume, A., Côté, G., \& Karimi, R. (2020). Scopus as a curated, high-quality bibliometric data source for academic research in quantitative science studies. \emph{Quantitative Science Studies}, \emph{1}(1), 377--386. \url{https://doi.org/10.1162/qss_a_00013}

\bibitem[\citeproctext]{ref-basu2012cognitive}
Basu, A., \& Wagner Dobler, R. (2012). {``Cognitive mobility''} or migration of authors between fields used in mapping a network of mathematics. \emph{Scientometrics}, \emph{91}(2), 353--368. \url{https://doi.org/10.1007/s11192-011-0613-5}

\bibitem[\citeproctext]{ref-bielick2018emergence}
Bielick, J., \& Laudel, G. (2018). The emergence of individual research programs in the early career phase of academics. \emph{Science, Technology, and Human Values}, \emph{43}(6). \url{https://doi.org/10.1177/0162243918763100}

\bibitem[\citeproctext]{ref-chakraborty2015understanding}
Chakraborty, T., Tammana, V., Ganguly, N., \& Mukherjee, A. (2015). Understanding and modeling diverse scientific careers of researchers. \emph{Journal of Informetrics}, \emph{9}(1), 69--78. \url{https://doi.org/10.1016/j.joi.2014.11.008}

\bibitem[\citeproctext]{ref-cohan2020specter}
Cohan, A., Feldman, S., Beltagy, I., Downey, D., \& Weld, D. (2020). {SPECTER}: Document-level representation learning using citation-informed transformers. \emph{{Proceedings of the 58th Annual Meeting of the Association for Computational Linguistics}}, 2270--2282. \url{https://doi.org/10.18653/v1/2020.acl-main.207}

\bibitem[\citeproctext]{ref-donner2020validation}
Donner, P. (2020). {A validation of coauthorship credit models with empirical data from the contributions of PhD candidates}. \emph{Quantitative Science Studies}, \emph{1}(2), 551--564. \url{https://doi.org/10.1162/qss_a_00048}

\bibitem[\citeproctext]{ref-enders2014turning}
Enders, J., Kehm, B. M., \& Schimank, U. (2014). Turning universities into actors on quasi-markets: How new public management reforms affect academic research. In \emph{The changing governance of higher education and research: Multilevel perspectives} (pp. 89--103). \url{https://doi.org/10.1007/978-3-319-09677-3_5}

\bibitem[\citeproctext]{ref-enduri2015does}
Enduri, M. K., Reddy, I. V., \& Jolad, S. (2015). {Does diversity of papers affect their citations? Evidence from American Physical Society Journals}. \emph{{11th International Conference on Signal-Image Technology \& Internet-Based Systems (SITIS)}}, 505--511. \url{https://doi.org/10.1109/SITIS.2015.60}

\bibitem[\citeproctext]{ref-fochler2018anticipatory}
Fochler, M., \& Sigl, L. (2018). Anticipatory uncertainty: How academic and industry researchers in the life sciences experience and manage the uncertainties of the research process differently. \emph{Science as Culture}, \emph{27}(3), 349--374. \url{https://doi.org/10.1080/09505431.2018.1485640}

\bibitem[\citeproctext]{ref-franzoni2017academic}
Franzoni, C., \& Rossi-Lamastra, C. (2017). Academic tenure, risk-taking and the diversification of scientific research. \emph{Industry and Innovation}, \emph{24}(7), 691--712. \url{https://doi.org/10.1080/13662716.2016.1264067}

\bibitem[\citeproctext]{ref-gieryn1978problem}
Gieryn, T. F. (1978). Problem retention and problem change in science. \emph{Sociological Inquiry}, \emph{48}(3/4), 96--115. \url{https://doi.org/10.1111/j.1475-682X.1978.tb00820.x}

\bibitem[\citeproctext]{ref-gingras2014criteria}
Gingras, Y. (2014). Criteria for evaluating indicators. In B. Cronin \& C. R. Sugimoto (Eds.), \emph{Beyond bibliometrics: Harnessing multidimensional indicators of scholarly impact} (pp. 109--125). Cambridge: MIT Press.

\bibitem[\citeproctext]{ref-glaser2015epistemic}
Gläser, J., Heinz, M., \& Havemann, F. (2015). Epistemic diversity as distribution of paper dissimilarities. In A. A. Salah, Y. Tonta, A. A. A. Salah, C. Sugimoto, \& U. Al (Eds.), \emph{{Proceedings of ISSI 2015 Istanbul: 15th International Society of Scientometrics and Informetrics Conference}}.

\bibitem[\citeproctext]{ref-glaser2015bibliometric}
Gläser, J., \& Laudel, G. (2015). A bibliometric reconstruction of research trails for qualitative investigations of scientific innovations. \emph{Historical Social Research/Historische Sozialforschung}, \emph{40}(3), 299--330. \url{https://doi.org/10.12759/hsr.40.2015.3.299-330}

\bibitem[\citeproctext]{ref-hagen2008harmonic}
Hagen, N. T. (2008). Harmonic allocation of authorship credit: Source-level correction of bibliometric bias assures accurate publication and citation analysis. \emph{PLoS One}, \emph{3}(12), e4021. \url{https://doi.org/10.1371/journal.pone.0004021}

\bibitem[\citeproctext]{ref-hargens1986migration}
Hargens, L. (1986). {Migration patterns of US Ph.D.s among disciplines and specialties}. \emph{Scientometrics}, \emph{9}(3-4), 145--164. \url{https://doi.org/10.1007/bf02017238}

\bibitem[\citeproctext]{ref-harmon1965profiles}
Harmon, L. R. (1965). \emph{{Profiles of Ph.D's in the Sciences: Summary Report on Follow-Up of Doctorate Cohorts, 1935-1960}}. \url{https://doi.org/10.17226/21462}

\bibitem[\citeproctext]{ref-hellsten2007self}
Hellsten, I., Lambiotte, R., Scharnhorst, A., \& Ausloos, M. (2007). Self-citations, co-authorships and keywords: A new approach to scientists' field mobility? \emph{Scientometrics}, \emph{72}(3), 469--486. \url{https://doi.org/10.1007/s11192-007-1680-5}

\bibitem[\citeproctext]{ref-hodge1981publication}
Hodge, S. E., \& Greenberg, D. A. (1981). Publication credit. \emph{Science}, \emph{213}(4511), 950--950. \url{https://doi.org/10.1126/science.213.4511.950.b}

\bibitem[\citeproctext]{ref-jia2017quantifying}
Jia, T., Wang, D., \& Szymanski, B. K. (2017). Quantifying patterns of research-interest evolution. \emph{Nature Human Behaviour}, \emph{1}(4), 0078. \url{https://doi.org/10.1038/s41562-017-0078}

\bibitem[\citeproctext]{ref-kawashima2015accuracy}
Kawashima, H., \& Tomizawa, H. (2015). {Accuracy evaluation of Scopus Author ID based on the largest funding database in Japan}. \emph{Scientometrics}, \emph{103}(3), 1061--1071. \url{https://doi.org/10.1007/s11192-015-1580-z}

\bibitem[\citeproctext]{ref-cetina1999epistemic}
Knorr Cetina, K. (1999). \emph{Epistemic cultures: How the sciences make knowledge}. Harvard University Press.

\bibitem[\citeproctext]{ref-lawson2014thematic}
Lawson, C., \& Soós, S. (2014). \emph{A thematic mobility measure for econometric analysis} (Working Paper No. 2014/2). {Library and Information Centre of the Hungarian Academy of Sciences, Department of Science Policy and Scientometrics}.

\bibitem[\citeproctext]{ref-lepair1980switching}
Le Pair, C. (1980). Switching between academic disciplines in universities in the {Netherlands}. \emph{Scientometrics}, \emph{2}, 177--191. \url{https://doi.org/10.1007/BF02016696}

\bibitem[\citeproctext]{ref-markus2010construct}
Markus, K. A., \& Lin, C. (2010). Construct validity. In N. J. Salkind (Ed.), \emph{Encyclopedia of research design} (pp. 229--233). Thousand Oaks, CA: SAGE Publications, Inc.

\bibitem[\citeproctext]{ref-muller2019re}
Müller, R., \& Kaltenbrunner, W. (2019). {Re-disciplining academic careers? Interdisciplinary practice and career development in a Swedish environmental sciences research center}. \emph{Minerva}, \emph{57}(4), 479--499. \url{https://doi.org/10.1007/s11024-019-09373-6}

\bibitem[\citeproctext]{ref-pramanik2019migration}
Pramanik, S., Gora, S. T., Sundaram, R., Ganguly, N., \& Mitra, B. (2019). On the migration of researchers across scientific domains. \emph{Proceedings of the international AAAI conference on web and social media}, \emph{13}, 381--392. \url{https://doi.org/10.1609/icwsm.v13i01.3238}

\bibitem[\citeproctext]{ref-rousseau2019knowledge}
Rousseau, R., Zhang, L., \& Hu, X. (2019). {Knowledge integration: Its meaning and measurement}. In W. Glänzel, H. F. Moed, U. Schmoch, \& M. Thelwall (Eds.), \emph{{Springer Handbook of Science and Technology Indicators}} (pp. 69--94). \url{https://doi.org/10.1007/978-3-030-02511-3_3}

\bibitem[\citeproctext]{ref-simonton2004creativity}
Simonton, D. K. (2004). \emph{Creativity in science: Chance, logic, genius, and zeitgeist}. Cambridge University Press.

\bibitem[\citeproctext]{ref-struck2025dependencies}
Struck, A. (2025). \emph{Dependencies in topic delineation and tracking} (PhD thesis, Humboldt-Universit{ä}t zu Berlin). \url{https://doi.org/10.18452/30669}

\bibitem[\citeproctext]{ref-tripodi2020knowledge}
Tripodi, G., Chiaromonte, F., \& Lillo, F. (2020). Knowledge and social relatedness shape research portfolio diversification. \emph{Scientific Reports}, \emph{10}(1), 14232. \url{https://doi.org/10.1038/s41598-020-71009-7}

\bibitem[\citeproctext]{ref-wang2020consistency}
Wang, Q., \& Schneider, J. W. (2020). Consistency and validity of interdisciplinarity measures. \emph{Quantitative Science Studies}, \emph{1}(1), 239--263. \url{https://doi.org/10.1162/qss_a_00011}

\bibitem[\citeproctext]{ref-yan2015understanding}
Yan, S., \& Lagoze, C. (2015). Understanding relationship between scholars' breadth of research and scientific impact. \emph{iConference 2015 Proceedings}.

\bibitem[\citeproctext]{ref-yu2021become}
Yu, X., Szymanski, B. K., \& Jia, T. (2021). Become a better you: Correlation between the change of research direction and the change of scientific performance. \emph{Journal of Informetrics}, \emph{15}(3), 101193. \url{https://doi.org/10.1016/j.joi.2021.101193}

\bibitem[\citeproctext]{ref-zeng2019increasing}
Zeng, A., Shen, Z., Zhou, J., Fan, Y., Di, Z., Wang, Y., \ldots{} Havlin, S. (2019). Increasing trend of scientists to switch between topics. \emph{Nature Communications}, \emph{10}, 3439. \url{https://doi.org/10.1038/s41467-019-11401-8}

\end{CSLReferences}
\end{document}